\documentclass[]{spie}  

\usepackage{amsmath,amsfonts,amssymb}
\usepackage{graphicx}
\usepackage[colorlinks=true, allcolors=blue]{hyperref}

\graphicspath{{../../images/}}
\usepackage{setspace}
\usepackage{float}
\usepackage[caption=false]{subfig}
\usepackage{wrapfig}
\usepackage{epstopdf}
\usepackage{slashbox}
\usepackage{tabularx}
\usepackage[numbered]{matlab-prettifier}
\usepackage{listings}
\usepackage[T1]{fontenc}
\usepackage{lmodern}

\def\bA{\mathbf{A}}
\def\bAhat{\mathbf{\hat{A}}}
\def\A{\mathcal{A}}

\def\th{\mathrm{th}}

\def \emis{\boldsymbol{\lambda}}

\def \sc {\mathrm{sc}}

\def \L {\mathcal{L}}

\def \P {\mathbb{P}}
\def \path{\mathbb{P}}

\def \pr{\mathrm{pr}}
\def \Pr{\mathrm{Pr}}

\def \obj {\boldsymbol{\lambda}}
\def\z{\boldsymbol{z}}

\title{Task-based assessment of binned and list-mode SPECT systems}

\author[1,2]{Md Ashequr Rahman}
\author[1,2]{Abhinav K. Jha}
\affil[1]{Department of Biomedical Engineering, Washington University in St. Louis, St. Louis, MO, USA}
\affil[2]{Mallinckrodt Institute of Radiology, Washington University in St. Louis, St. Louis, MO, USA}

\authorinfo{Further author information: (Send correspondence to Abhinav K. Jha)\\Abhinav K. Jha: E-mail: a.jha@wustl.edu}

\begin{document} 
\noindent This manuscript has been accepted to SPIE Medical Imaging, February 15-19, 2021. Please use the following reference when citing the manuscript.\\

\noindent Rahman, M. A. and Jha, A. K., ``Task-based assessment of binned and list-mode SPECT systems," Proc. SPIE Medical Imaging, 2021.
\clearpage
\maketitle

\begin{abstract}
In SPECT, list-mode (LM) format allows storing data at higher precision compared to binned data. 
There is significant interest in investigating whether this higher precision translates to improved performance on clinical tasks. Towards this goal, in this study, we quantitatively investigated whether processing data in LM format, and in particular, the energy attribute of the detected photon, provides improved performance on the task of absolute quantification of region-of-interest (ROI) uptake in comparison to processing the data in binned format. 
We conducted this evaluation study using a DaTscan brain SPECT acquisition protocol, conducted in the context of imaging patients with Parkinson's disease. This study was conducted with a synthetic phantom. 
A signal-known exactly/background-known-statistically (SKE/BKS) setup was considered. 
An ordered-subset expectation-maximization algorithm was used to reconstruct images from  data acquired in LM format, including the scatter-window data, and including the energy attribute of each LM event. Using a realistic 2-D SPECT system simulation, quantification tasks were performed on the reconstructed images. 
The results demonstrated improved quantification performance when LM data was used compared to binning the attributes in all the conducted evaluation studies. Overall, we observed that LM data, including the energy attribute, yielded improved  performance on absolute quantification tasks compared to binned data. 
\end{abstract}


\section{INTRODUCTION}
\label{sec:intro}  

Single photon emission computed tomography (SPECT) is a widely used clinical imaging modality with multiple clinical applications. In SPECT, using the detector measurements, for each detected photon, multiple attributes including the position of interaction in the detector, energy deposited in the detector and angle of acquisition can be estimated and stored. Recording attributes for each detected photon results in list-mode (LM) data format \cite{Barrett_1997}. Traditional methods typically bin the attributes before reconstructing the image. 
Studies have shown that this process of binning leads to information loss \cite{caucci2019towards, Jha:15:pmb, clarkson2020quantifying, rahman2020fisher}. 
In particular, it  was shown that a larger set of null functions exist when binning is introduced \cite{Jha:15:pmb, caucci2013image, clarkson2020quantifying}. The increase in null function results in bias on the task of quantification \cite{Jha:15:spie}.
Further, it has also been shown in a simplified SPECT system that the process of binning the angle of acquisition attribute negatively impacts performance on the task of quantification \cite{Jha:15:spie}.
These findings motivate studying the impact of binning on quantification performance for more realistic SPECT systems. 

One key attribute estimated by a SPECT system for each gamma-ray photon is the energy deposited by the photon in the scintillation crystal. Studies are suggesting that this energy attribute contains information to perform tasks such as estimating the  attenuation distribution \cite{rahman2020fisher, Jha:13:spie, Cade_2013, zitong}. However, typically this attribute is binned, leading to different energy windows, and of these, the scatter-window data is typically discarded. More recent studies are demonstrating that photons in the scatter-window data may contain information to estimate the activity distribution \cite{rahman2020fisher, kadrmas1997analysis}. Thus, that leads to the question of whether processing these photons in LM format without binning the energy attribute results in improved image quality.  
Our goal in this manuscript was to investigate this question. This would provide further insights on the impact of processing binned vs. LM data on image quality in SPECT. 

Imaging systems and methods are developed for specific clinical tasks, such as detection, quantification, or a combination of both. Thus, a rigorous approach to image-quality evaluation should account for the clinical task performed on the acquired image. Objective assessment of image quality (OAIQ)-based studies provide a mechanism for conducting such task-based evaluation quantitatively and by accounting for population variability, imaging-system physics, the observer performing the task, and a figure of merit that quantifies this performance \cite{B&M, barrett1993model, frey2002application, jha2013ideal, li2016use, yu2020ai, marin2013numerical,jha2016no}. 
Thus, we conducted an OAIQ-based study to evaluate the impact of binning on performance in quantification tasks. 

A key enabler of our OAIQ-based study was a recently developed iterative method that uses LM data, including scatter-window data and containing the energy attribute, to reconstruct the activity distribution in SPECT \cite{rahman2020list}. This method compensates for the relevant image-degrading processes in SPECT including scatter, attenuation, noise, and collimator-detector response. The method provides an avenue to investigate the impact of binning the LM event attributes, including the energy attribute, on performance in absolute quantification tasks in SPECT. Our objective in this study was to investigate this impact using an OAIQ framework on the task of absolute quantification in SPECT. For this purpose, we conducted OAIQ-based studies in the context of quantiative dopamine transporter (DaT) SPECT.
Quantitative uptakes in striatal regions in the brain, as extracted for DaT SPECT images, are being explored as biomarkers to measure severity of Parkinson's disease. There is an important need for such biomarkers, given the exciting developments in developing disease-modifying therapies for this disease. However, for these quantitative uptakes to be applicable as biomarkers, they need to be reliably estimated from the images. 
We investigated whether using LM data would yield more reliable quantification of these uptake values. 

\section{Method}
A simulation-based OAIQ study typically consists of five components: (a) a definition of the task, (b) object model, (c) simulation of the imaging process, (d) process to extract the task-specific information and (e) figure of merit to measure task performance. In this section, we describe each of these components in our study.  As mentioned above, the study was conducted in the context of quantitative DaT-SPECT. 
\begin{figure}
\centering

\subfloat[]{
\includegraphics[height = 1.2in]{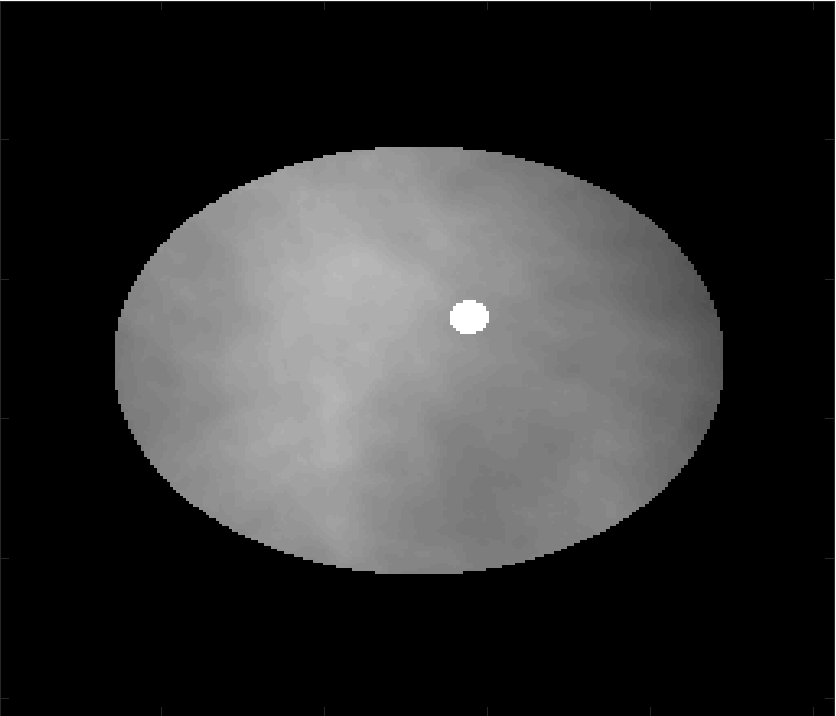}
\label{fig:synth_true_act}
}
\subfloat[]{
\includegraphics[height = 1.2in]{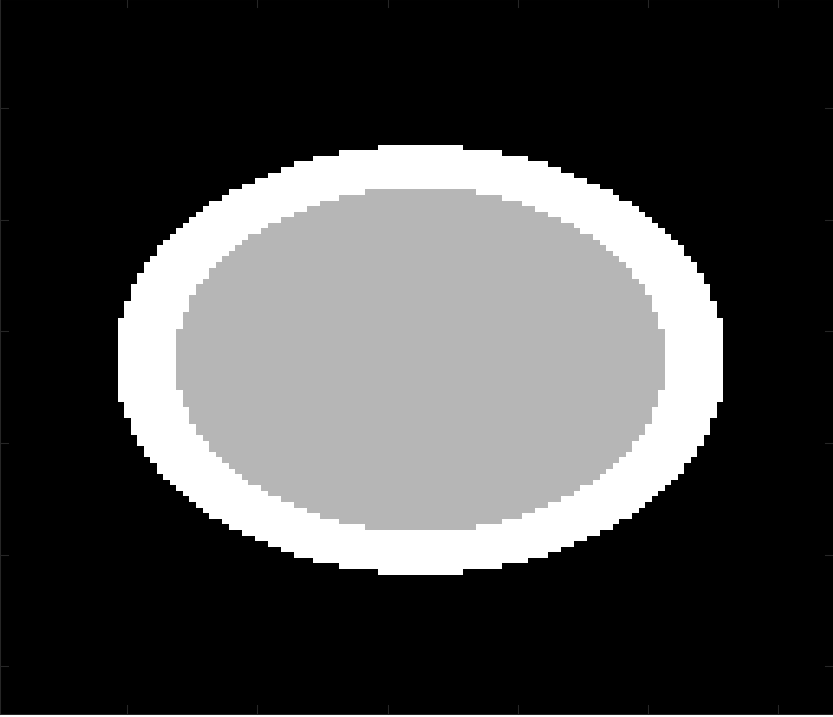}
\label{fig:synth_true_attn}
}

\caption{For synthetic phantom, (a) a realization of true activity distribution for signal-present case where mean SBR = 2:1 and (b) true attenuation distribution.}
\label{fig:all_true_act_attn}
\end{figure}
\subsection{Definition of task}
We evaluated the LM and binned systems on the task of quantifying activity within a pre-defined region of interest (ROI). In DaT-SPECT, the goal is to estimate the activity in the caudate and putamen regions. The boundaries for these regions can be obtained by either segmenting these regions \cite{boudraa2006image, moon2020physics, liu2021fully}, or from other imaging modalities, such as MRI \cite{colloby2004application}. Thus, the ROI can be defined. We thus frame our task as estimating the activity within the known ROIs from the SPECT images. These SPECT images were acquired using a 2D acquisition protocol, as described in more detail in the next section. 

We conducted the quantitative analysis with a synthetic phantom that has similar properties as the brain phantom. Task performance was assessed over multiple realizations. The background was also varied in the different realizations, while the signal properties were constant. Thus, this was a signal-known-exactly/background-known-statistically (SKE/BKS) study in the context of a quantification task except that the signal intensity was not known. The rationale behind conducting this study was to evaluate performance as the background varied on quantification tasks. Variation in background has been known to impact detection-task performance \cite{rolland1992effect}.

\subsection{Object model}
To simulate an object similar to the brain, an elliptical shaped synthetic phantom (Fig.~\ref{fig:synth_true_act}) was considered where the background was simulated using a clustered lumpy background model \cite{bochud1999statistical}. A circular region with $7$ mm radius located at an off-center position inside the phantom was considered as the known signal. To analyze the effect of varying signal/tumor-to-background ratio (SBR/TBR) on task performance, the mean SBR value was varied between 1.2:1, 1.5:1, 2:1 and 3:1. The attenuation distribution was considered to be non-uniform (Fig.~\ref{fig:synth_true_attn}). The attenuation coefficient was $21$m$^{-1}$ at the rim region of phantom  which simulates a skull and  $15$m$^{-1}$ at the inner region simulating the soft tissue. 

\subsection{Simulating the imaging process}
We simulated a 2-D SPECT system with configuration similar to
GE's Optima 640 system and a low-energy high resolution (LEHR) parallel-hole collimator. We considered that the phantom has been injected with Ioflupane (I-123) radiotracer, which emits gamma-ray photon at 159 KeV energy. The detector position resolution was set to 4~mm. The detector energy resolution was set to $10\%$ FWHM at 159 KeV and was assumed to be constant in the range of 80 KeV to 175 KeV. LM data, that falls into the energy range 70 KeV to 175 keV, was generated using Monte Carlo simulation and only contained events that went through up to first order of scatter. The data was acquired in 120 fixed angular position over $360^{\circ}$. Approximately $8 \times 10^4$ events were detected in a energy window ranging from 80 KeV to 175 KeV, thus simulating a low-dose SPECT protocol.

This LM data was reconstructed as such using the reconstruction algorithm described below. To study the impact of binning on task performance, we binned the energy attribute of this data into multiple bins, namely 2 bins and 3 bins. We set the first bin as same as the photo-peak energy
window (143-175 KeV for DaT-SPECT). All the events inside this bin were assigned a fixed energy value, equal to the primary photon energy (159 KeV). Depending on the total number of
bins, the scatter-window was then divided into equal-width windows. All photons within a given scatter-energy bin were assigned the energy value at the center of that bin. These binned data were also reconstructed using the reconstruction algorithm described below. 
\subsection{Process to extract task-specific information }
The measured projection data was reconstructed using the maximum-likelihood expectation-maximization-based method as briefly described in Rahman et al. \cite{rahman2020list}. The method extends upon an approach originally proposed in \cite{Jha:13:spie,Jha:13:diss} and allows processing LM data acquired in any arbitrary-sized energy window and incorporate the energy attribute of detected photon. 
Further, the method compensates for attenuation, scatter, collimator-detector response, and noise in SPECT.
Here we briefly describe the technique. 

Consider a SPECT system that acquires LM data for a fixed scan time, T. The objective of the reconstruction method is to estimate the discrete activity distribution denoted by the $N$-dimensional vector $\emis$ over $Q$ voxels. Let $\lambda_q$ denote the mean rate of emission from $q^\text{th}$ voxel. 
Denote the number of detected events by $J$.
The detected events are stored in LM format where for each photon, the attributes of position of interaction of detected photon in the the scintillation crystal, energy deposited by the detected photon and time of detection are recorded. We denote the true and measured attributes of $j^\th$ detected event by vectors $\bA_j$ and $\bAhat_j$, respectively. We also denote $\P$ as the path traversed by an emitted photon and $\hat{\A}$ as the full set of measured attributes where $\hat{\A} = \{\bAhat_j,j=1,2,\ldots J\}$.
We can write the likelihood of the measured LM data as
\begin{align}
\pr(\hat{\A},J|\obj) &= \Pr(J|\obj)\pr(\hat{\A}|\obj)\nonumber\\
 &= \Pr(J|\obj)\prod_{j=1}^{J}\pr(\bAhat_j|\obj)\\
 &=\Pr(J|\obj)\prod_{j=1}^{J}\sum_{\P}\pr(\bAhat_j|\P)\Pr(\P|\obj),
\label{eq:lm_ll_gen}
\end{align} 
where, in the second step, we have used the fact that the LM events are all independent, while in the third step, we have decomposed $\pr(\bAhat_j | \obj)$ as a mixture model in terms of the probability that a photon takes a path and the probability of that path. Here, we point that $J$ is a Poisson distributed random variable. 
Let $s_{eff}(\P)$  denote the sensitivity of a path $\P$ and $\lambda(\P)$ denote the activity at the voxel where the path originates from. We can derive the form of  $\pr(\bAhat_j|\P)$ using radiative transport equation as \cite{Jha:13:diss}
\begin{align}
\pr(\P|\obj)=\frac{\lambda({\P})s_{eff}(\P)}{\sum_{\P'}\lambda({\P'})s_{eff}(\P')}.
\label{eq:pr_path_1}
\end{align}
Thus we can write the log-likelihood of observed LM data as
\begin{align}
\L(\emis|\hat{\A},J)=\sum_{j=1}^{J}\log\sum_{\P}\pr(\bAhat_j|\P)\lambda(\P)s_{eff}(\P) + J\log( T)-T\sum_{\P}\lambda(\P)s_{eff}(\P) - \log J!.
\label{eq:lll_med}
\end{align}
To maximize this likelihood, an expectation-maximization algorithm is developed. For each event $j$ and each possible path $\P$, we denote a hidden variable $z_{j,\P}$ where 
\[
 z_{j,\P} = 
  \begin{cases} 
   1 & \text{if event $j$ took the path $\P$}. \\
   0       & \text{otherwise}.
  \end{cases}
\]
Let us denote the vector $\z_{j}$ by accumulating all the hidden vectors of all possible path for this specific event $j$. The observed LM data and the hidden vectors form the complete data. We can derive the form of complete data log-likelihood as:
\begin{align}
&\L_C(\emis|\{\hat{A_j},\z_j;j=1,\ldots J\},J)\nonumber\\
&=\sum_{j=1}^{J}\left[\sum_{\P}z_{j,\P}\left\{\log \pr(\hat{\A_j}|\P)+\log \lambda({\P}) + \log s_{eff}(\P)\right\}\right]  + J \log T-T\sum_{\P}\lambda({\P})s_{eff}(\P) - \log J!.
\label{eq:cd_lll}
\end{align}
Taking the expectation and maximization steps on this complete data log-likelihood leads to following iterative update:
\begin{align}
\hat{\lambda}_q^{(t+1)}=\dfrac{\sum_{j=1}^{J}\sum_{\P_q}\bar{z}_{j,\P_q}^{(t+1)}}{T\sum_{\P_q}s_{eff}(\P_q)},
\label{eq:lm_mlem}
\end{align}
where, $\P_q$ denotes paths that start from voxel $q$, $\lambda_q$ denotes the activity rate at that starting voxel and  $\bar{z}_{j,\P}^{(t+1)}=E\left[z_{j,\P}|\bAhat_j;\emis^{(t)}\right]$ is the expected value of the hidden variable conditioned on observed LM data. 

However, the algorithm was still computationally intensive. Thus, in our OAIQ-based studies, we considered only up to first-order scatter events. Further, to ensure that the number of scatter events in the forward and inverse model match, we reconstructed the images using photons that scatter only once.

From the reconstructed images, we estimated the mean activity uptake inside the known region of interest (ROI). 
Let the $N$-dimensional vector $\boldsymbol{\chi}_k$ denote the binary mask that denotes the ROI for the $k^{\th}$ region. Let $\hat{a}_k$ denote the estimated activity in that region using this procedure. Then, mathematically, 
\begin{align}
\hat{a}_k  = \frac{\sum_{q=1}^{Q}\lambda_{q} \chi_{qk}} {\sum_{q=1}^{Q}\chi_{qk}}.
\end{align}

\subsection{Figure of merit to measure task performance}
The performance of the list-mode and binned-based estimation approaches on the task of activity estimation was quantified on the basis of accuracy and precision of the estimated uptake using the metrics of normalized bias and variance. Further, the overall reliability of the estimated uptake was quantified using the metric of normalized root mean squared error (NRMSE). This is a summary figure of merit that quantifies the effect of both bias and variance, and thus indicates the overall reliability on the quantification task. Let $a_k$ and $\hat{a}_k$ denote the true and estimated activities within the $k^{\th}$ ROIs.  Let $P$ denote the number of noise realizations. Then, the NRMSE for the $k^{\th}$ ROI was defined as:
\begin{align}
NRMSE = \dfrac{1}{{{a}}_k}\sqrt{\dfrac{\sum_{i=1}^{P}(a_k-\hat{a}_{ki})^2}{P}}.
\end{align}

\section{Results}

\begin{figure}[h!]
\centering

\subfloat[]{
\includegraphics[width = 2.8 in]{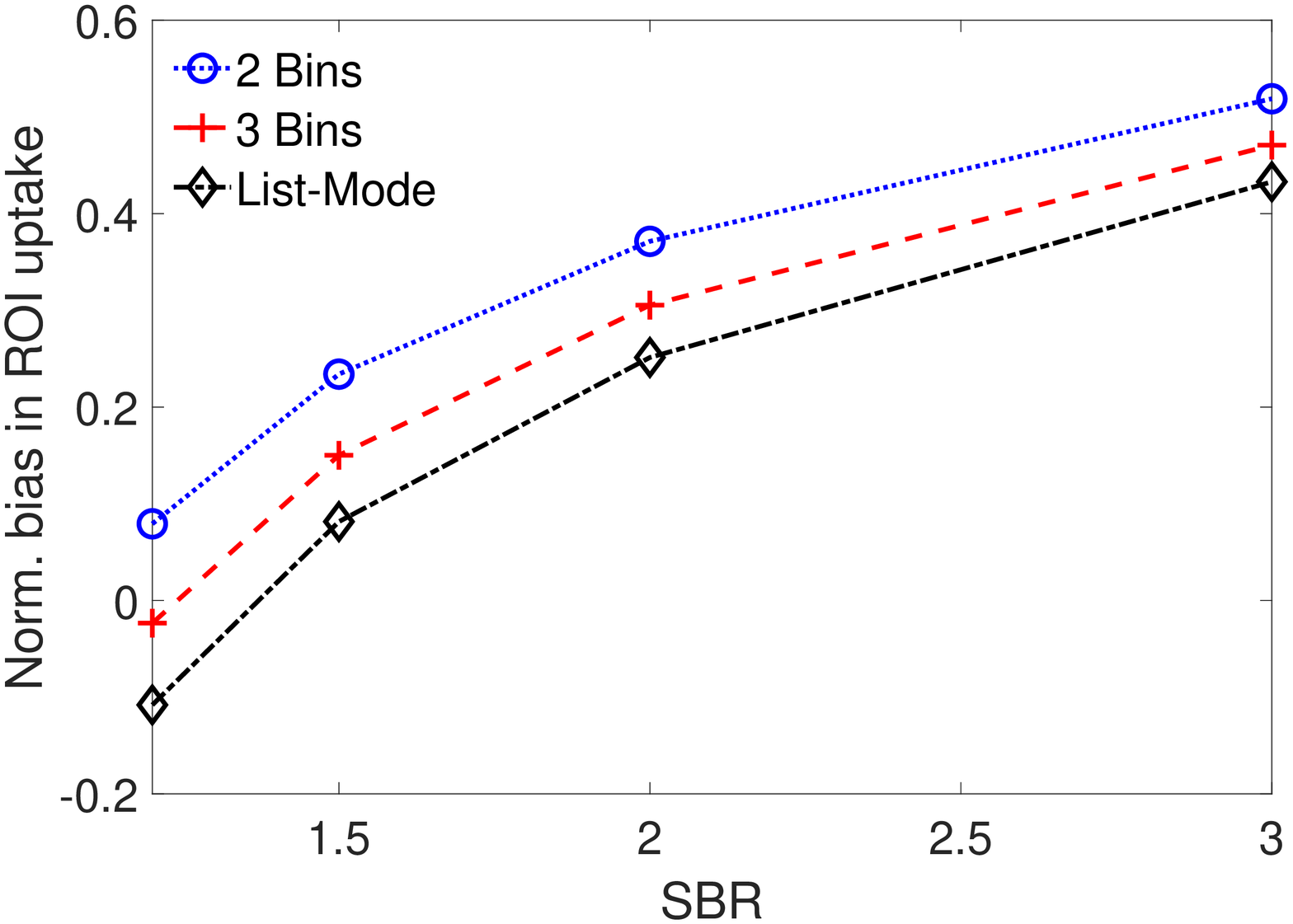}
\label{fig:Bias_vs_TBR}
}
\subfloat[]{
\includegraphics[width = 2.8 in]{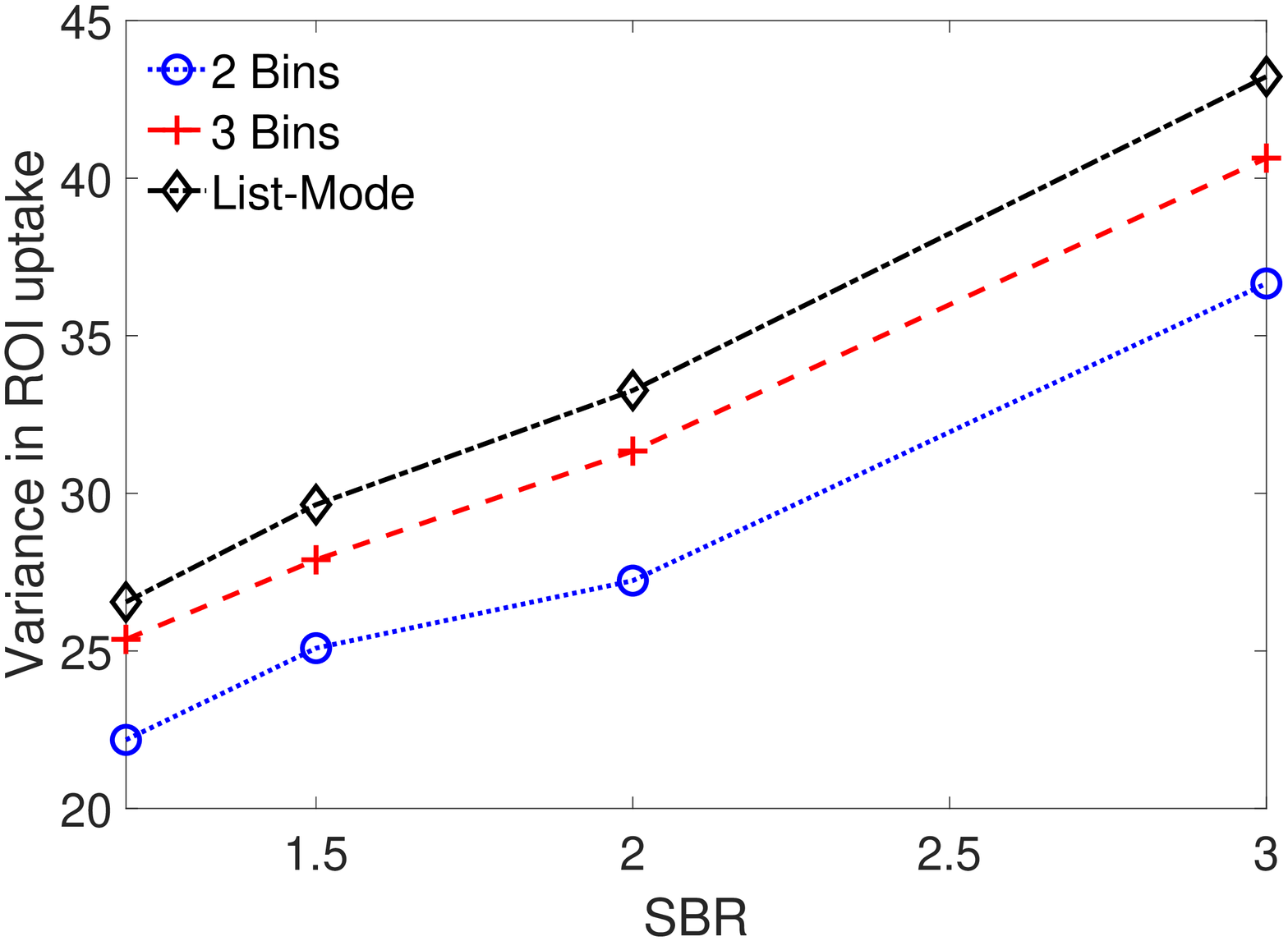}
\label{fig:Var_vs_TBR}
}
\hfill
\subfloat[]{
\includegraphics[width = 2.8 in]{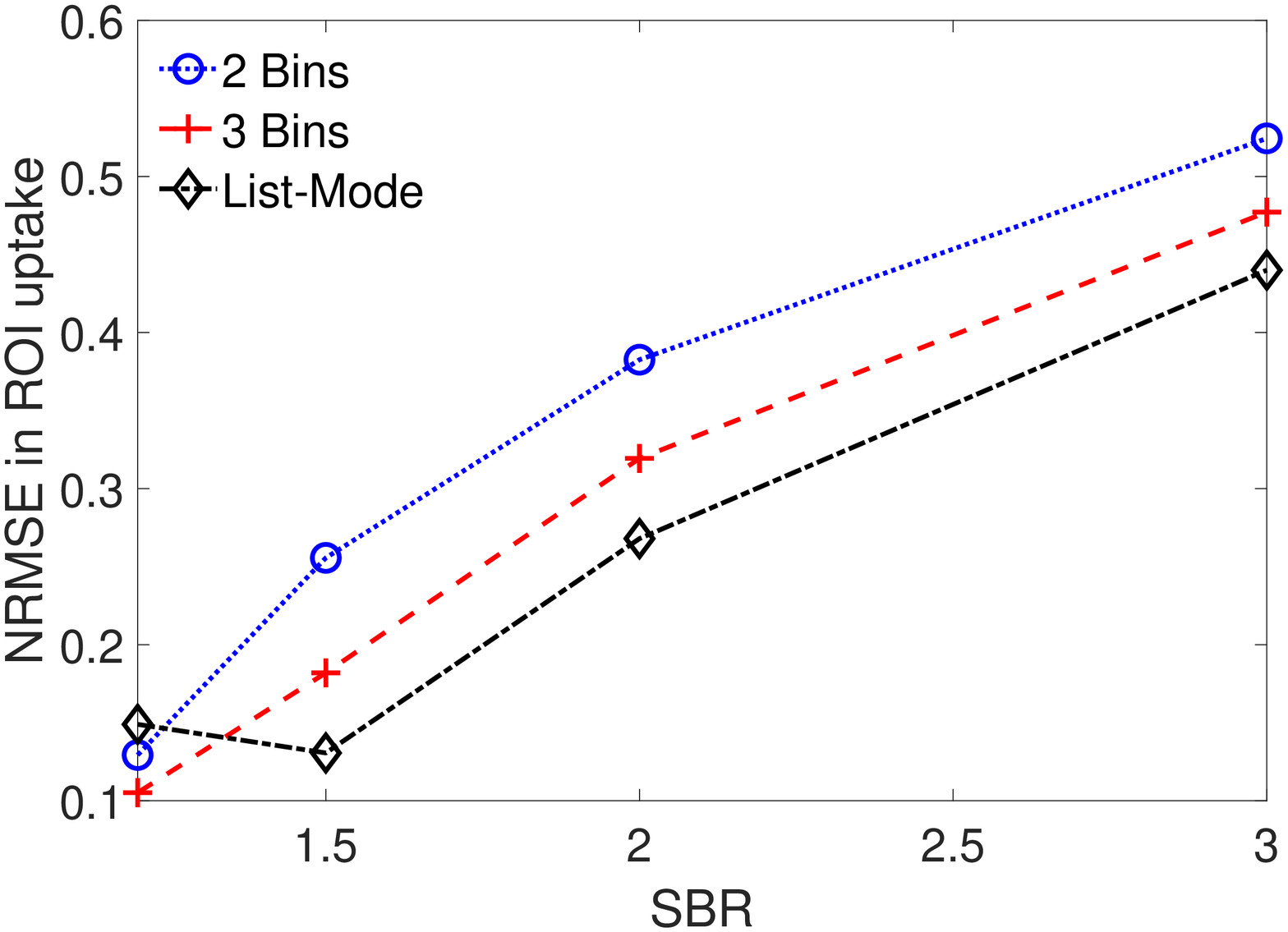}
\label{fig:NRMSE_vs_TBR}
}
\caption{For different binning configuration,  (a) normalized bias, (b) variance and (c) NRMSE between the estimated activity uptake inside signal region and true mean activity uptake as a function of mean SBR in a SKE/BKS setup.}

\end{figure}

Fig.~\ref{fig:Bias_vs_TBR} shows that processing the energy attribute in LM format led to a lower bias for SBR values greater than 1.5 in comparison to binning this attribute. Binning the list-mode attributes leads to increase in null space \cite{Jha:15:pmb, clarkson2020quantifying}, and as shown in another study \cite{Jha:15:spie}, this causes a bias in quantification performance. Our results are consistent with that study and show that binning the energy attribute leads to an increase in bias. 

At the same time, we observe that the standard deviation when using the LM data is higher compared to when using the binned format (Fig.~\ref{fig:Var_vs_TBR}). However, in case of the NRMSE, which is a combination of the bias and variance, it is observed that LM data yields a lower NRMSE compared to the binned format (Fig.~\ref{fig:NRMSE_vs_TBR}). This suggest that the improvement in bias is more significant compared to the impact on increasing variance when using list-mode data

\section{Discussions}
 In this manuscript we evaluated the effect of binning the energy attributes of the LM data on the task of quantifying uptake within a pre-defined ROI in a 2-D DaT-SPECT system. We consistently observed that processing data in LM format yielded improved quantification performance. 
As mentioned above, LM format allows storing the data at higher precision compared to binned format.  
Our results show that this increase in precision very much translates into a tangible improved performance on quantification tasks. 
 
In our quantification study, we did not perform any partial volume compensation (PVC). Previous studies \cite{1490666} have demonstrated that compensating for PVEs results in improved quantification performance. Thus, integrating this LM-based reconstruction approach with a PVC approach may lead to even more reliable quantification, and is an important area of future research.  

The current study has some limitations. We evaluated the performance of LM and binned data based on a 2-D SPECT system. We considered only up to single-order of scatter in forward model and reconstruction framework. Our promising results motivate extension of this study to 3-D setup and modeling multiple scatter events. 

\section{Conclusions}
Our objective-assessment-of-image-quality-based analysis provides evidence that processing SPECT-emission data in list-mode format provides improved performance compared to binned format for the task of estimating activity within a known region of interest, as evaluated in clinically relevant applications conducted in the context of  quantitaitve dopamine transporter SPECT. This analysis motivates large-scale 3D simulation and physical-phantom studies to further validate this finding. Improved performance with list-mode data in these studies would provide evidence to process the energy attribute in list-mode format as opposed to binning data in energy windows.

\appendix    

\acknowledgments 
This work was financially supported by NIH R21 EB024647
(Trailblazer award) and by an NVIDIA GPU grant. The authors thank Richard Laforest, PhD for helpful discussions.
\bibliography{refs_final} 
\bibliographystyle{spiebib} 

\end{document}